\documentclass[12pt]{article}

\usepackage[pdftex]{graphicx}
\usepackage[absolute]{textpos}

\newcommand{\dpar}      [2] {\frac{\partial #1}{\partial #2}}

\newcommand{\ndpar}     [3] {\frac{\partial^{#1} #2}{\partial #3^{#1}}}
\newcommand{\schr}{Schr\"o\-din\-ger}
\newcommand{\nlse}{nonlinear \schr\ equation}

\newcommand{\ivp}{initial-value problem}
\newcommand{\ist}{inverse scattering transform}
\newcommand{\ibvp}{initial-boundary value problem}
\newcommand{\D}{Dirichlet}
\newcommand{\N}{Neumann}
\newcommand{\R}{Robin}
\newcommand{\bc}{boundary condition}

\renewcommand{\P}{periodic}
\newcommand{\eNss}{exact $N$-soliton solution}

\newcommand{\s}{soliton}

\newcommand{\text}[1]{\mbox{#1}}
\newcommand{\sech}{\text {\,sech}}
\newcommand{\eqref}[1]{(\ref{#1})}
\newcommand{\im}{{\cal I}m}
\newcommand{\re}{{\cal R}e}

\newcommand{\Z}{\bf Z}
\title{A Quantum Mechanics Analogy for the Nonlinear Schr\"{o}dinger Equation
in the Finite Line}

\author{J. I. Ramos and F. R. Villatoro \\
    Departamento de Lenguajes y Ciencias de la Computaci\'on \\
    E. T. S. Ingenieros Industriales \\
    Universidad de M\'alaga \\
    Plaza El Ejido, s/n \\
    29013-M\'alaga  \\
    SPAIN \\
}

\date{}

\begin{document}

\maketitle

\begin{abstract}

A quantum mechanics analogy is used to determine the forces acting on
and the energies of solitons governed by the \nlse\ in  finite intervals with
periodic and with homogeneous Dirichlet, Neumann and Robin boundary conditions.
It is shown that the energy densities remain nearly constant for periodic, while
 they undergo large variations for homogeneous, boundary conditions. The largest
variations in the force and energy densities occur for the Neumann boundary
conditions, but, for all the boundary conditions considered in this paper, the
magnitudes of these forces and energies recover their values prior to the
interaction of the soliton with the boundary, after the soliton rebound process
is completed. It is also shown that the quantum momentum changes sign but
recovers its original value after the collision of the soliton with the
boundaries. The asymmetry of the Robin  boundary  conditions shows different
dynamic behaviour at the left and right boundaries of the finite interval.

\vspace{1cm}

KEYWORDS: Nonlinear Schr\"{o}dinger equation, two-point initial-value problems,
quantum analogy, forces

\end{abstract}

\section{Introduction}

\label{se:introduction}

The one-dimensional \nlse\ is ubiquitous in many branches of mathematics, physics
and engineering~\cite{Ablo81} and ~\cite{Lamb80}, e.g., fluid mechanics, plasma
physics, nonlinear optics, nonlinear wave propagation, etc. Its \ivp\ is
integrable by the \ist\ and has a truly remarkable analytical structure.
Despite the great interest that this equation has received in the past two
decades through  analytical and numerical studies, the existence and uniqueness
of the \nlse\  in the quarterplane, i.e., for semi-infinite spatial domains,
seems to remain an issue of current research~\cite{Chu90}.

Three main lines of research on the \nlse\ in the quarterplane have been
undertaken. The first is based on results obtained numerically and uses ad hoc
assumptions~\cite{Kaup86}. The second one is based on the use of the inverse
scattering transform and B\"acklund transformations
~\cite{Bikb91}, while the third one employs the infinite
sine Fourier transform~\cite{Foka89}. However, these three lines of research
have been concerned with the generation of solitons at the boundaries, e.g.,
boundary-generated solitons.

Ablowitz and Segur~\cite{Ablo75} showed that the quarter-plane, \ibvp\ is
solvable by means of the nonlinear, sine Fourier transform if certain symmetry
conditions hold. In particular, Ablowitz and Segur~\cite{Ablo75} proved the
solvability of semi-infinite line problems subject to homogeneous Dirichlet or
Neumann conditions. Fokas~\cite{Foka89} and
Bikbaev and  Tarasov~\cite{Bikb91} have extended this result to
semi-infinite line problems subject to homogeneous Robin boundary
conditions. Fokas~\cite{Foka89}
has studied the generation of solitons by a non-homogeneous  \R\ \bc\ in the
semi-infinite line with zero initial conditions.

Some authors believe in  the almost integrability of the \nlse\ in the
quarterplane, also called `forced integrable
\nlse'~\cite{Kaup85b} because the \bc s can be viewed as forces that act at
the boundaries. Kaup~\cite{Kaup85b} showed that the difficulty in the
solvability of the forced \D\ initial-boundary value problem in the
quarterplane  lies in the  determination of the spatial derivative of the
amplitude at the boundary, and in  the determination of its primitive in the \N\
case. However, his studies have  not been able to provide   a connection, if
there is any, between  solitons in initial-value and  quarterplane problems.
Other authors, such as Kaup~\cite{Kaup85a} and  Kaup and
Hansen~\cite{Kaup86}  have  studied both
analytically and numerically the generation of solitons in quarter-plane
problems
subject to non-homogeneous Dirichlet boundary conditions.

If little is known about the existence and uniqueness of solutions of the \nlse\
in the quarterplane, much less is known about that equation in finite intervals,
i.e., two-point, initial-boundary value problems, because neither the inverse
scattering transform nor the infinite sine Fourier transform can be applied to
them. Numerical solutions, however, may be obtained for these problems by
employing, for example, global spectral, finite difference or finite element
methods. In this paper, a numerical study of the \nlse\ subject to periodic and
to homogeneous Dirichlet, Neumann and Robin boundary conditions in finite
intervals is presented in order to determine the forces on and the energies of
the solitons by using a quantum mechanics analogy. An analogy similar to the one
presented in this paper has been previously used by Bian and Chan~\cite{Bian91}
who only studied the nonlinear and diffraction forces on solitons governed by the
initial-value problem of the \nlse. These authors, however, used a filter and
their force and energy densities are not consistent with a strict quantum
mechanics analogy.

\section{The One-Dimensional, Nonlinear Schr\"{o}din\-ger
         E\-quation }

\label{se:forces}

The one-dimensional, time-dependent, \nlse\  with a cubic
potential can be written as
    \begin{equation} \label{eq:qnse}
i \hbar\ \hat{u}_{\hat{t}} = -\frac {\hbar^2}{2 m} \hat{u}_{\hat{x}\hat{x}}
               + V \hat{u}
    \end{equation}
where the potential well has the following form
    \begin{equation} \label{eq:pot}
V = -q \frac {\hbar^2}{m} | \hat{u} |^2
    \end{equation}
$\hat{t}$ is time, $\hat{x}$ is the spatial coordinate, $\hat{u}$ is a complex
amplitude, $\hbar$ is the Planck constant, $q$ is a coefficient, $m$ is the
mass of the soliton considered as a quantum particle, and the subscripts denote
partial differentiation.

A quantum-like analog of the \nlse\ can be easily obtained by assuming that
the  complex amplitude, $\hat{u}$, of the \nlse\ is like the wave function of
a quantum particle. This allows to determine its quantum momentum, quantum
energy and forces as follows. The linear quantum momentum is defined by the
one-dimensional operator~\cite{Land77}
    \begin{equation} \label{eq:mom:op}
\hat{P} \equiv - i \hbar \dpar{}{\hat{x}}
    \end{equation}
whose mean value is given by the following time-dependent expression
    \begin{equation} \label{eq:mom:mv}
\langle P \rangle = \frac { \langle \hat{u},\hat{P} \hat{u} \rangle } {\langle
 \hat{u}, \hat{u}
\rangle}
  = - i \hbar \frac { \int_{\cal D} \hat{u}^* \hat{u}_{\hat{x}} \,d{\hat{x}} }
            {\int_{\cal D} \hat{u}^* \hat{u}\,d{\hat{x}} } 
    \end{equation}
where $\cal D$ is the spatial domain, and $\hat{u}^*$ denotes the complex
conjugate of $\hat{u}$.

The quantum
momentum density may be defined as
    \begin{equation} \label{eq:mom:ld}
p(\hat{x},\hat{t}) = -i \hbar {\hat{u}^* \hat{u}_{\hat{x}}}.
    \end{equation}
The linear quantum energy is defined by the operator
    \begin{equation} \label{eq:ene:op}
\hat{E} = i \hbar \dpar{}{{\hat{t}}}
    \end{equation}
whose mean value is
    \begin{equation} \label{eq:ene:mv}
\langle E \rangle = \frac { \langle \hat{u},\hat{E} \hat{u} \rangle }
              {\langle \hat{u}, \hat{u} \rangle}
  =  i \hbar \frac { \int_{\cal D} \hat{u}^* \hat{u}_{\hat{t}} \,d{\hat{x}} }
           {\int_{\cal D} \hat{u}^* \hat{u}\,d{\hat{x}} }.
    \end{equation}
The  energy density is
    \begin{equation} \label{eq:ene:ld}
e(\hat{x},\hat{t}) = i \hbar  {\hat{u}^* \hat{u}_{\hat{t}}}.
    \end{equation}

Equation \eqref{eq:qnse} can be nondimensionalized using the linear
transformations
    \begin{equation} \label{eq:scale}
\hat{u} \rightarrow \sqrt{\frac {m}{q \hbar} } u , \qquad
\hat{t} \rightarrow t, \qquad
\hat{x} \rightarrow \sqrt {\frac {\hbar}{2 m} } x
    \end{equation}
to yield
    \begin{equation} \label{eq:nse}
i u_{t} = -u_{xx} - | u |^2 u
    \end{equation}

Hereon, unless otherwise stated, nondimensional
variables will be used. Note that the above dimensional definitions of the
quantum momentum and energy  may be used to obtain their
dimensionless counterparts by simply setting $\hbar$ and $m$ equal to
unity in the definitions of the dimensional quantum momentum and energy
densities.

The energy density can be written  as
    \begin{equation} \label{eq:lde:comp}
e(x,t) = -( u^*u_{xx} + |u|^4) = e_k(x,t) + e_v(x,t)
    \end{equation}
where the  kinetic and potential energy densities are, respectively,
    \begin{equation}
e_k(x,t) = - u^*u_{xx}, \qquad
e_v(x,t) = - |u|^4.
\end{equation}

The spatial derivatives of the energy densities define effective force
densities as follows. From the potential energy density, the following nonlinear
force density that produces  self-focusing on solitons is obtained
    \begin{equation} \label{eq:lde:ev2}
f_n(x,t) = - \dpar{}{x} e_v(x,t)
  = \dpar{}{x} \left( {|u|^4} \right)
    \end{equation}
while, from the kinetic one, the following diffraction force density that is
responsible  for the diffraction effect on the \s\ results
    \begin{equation} \label{eq:lde:ev3}
f_d(x,t) = - \dpar{}{x} e_k(x,t)
  = \dpar{}{x} \left(  {u^*u_{xx}} \right) .
    \end{equation}

The kinetic energy and diffraction force densities defined above have complex
values. In quantum mechanics, only the mean values of the energy and
force can be  physically measured and these values are always real due to the
Hermitian quantum  operators.

In order to obtain physical insight  from the energy and force densities
defined above, it is  convenient to use real values by taking either their real
or  imaginary part.  In this paper, the
real parts of the energy and  force densities are used.

A physically and mathematically consistent use of the quantum
analogy may also be obtained by defining the energy density as
$e/\varrho$ where $\varrho$ is a constant. For example, the following
renormalization factor may be used for the initial value problem of the \nlse
\begin{equation} 
\varrho = {\int_{\cal D} u^* u\,dx }
\end{equation}
which coincides with the first invariant of the \ivp\ of the \nlse. However,
this value of $\varrho$ may not be constant for two-point,
initial-boundary-value problems (cf. Section 4), i.e., it may not be used to
obtain a physically and mathematically consistent quantum analogy for two-point,
initial value problems.

Bian and
Chan~\cite{Bian91} defined the energy density as $e/\varrho$ where
$\varrho$=$u^*u$ in order to assess  physically  that a soliton can be
considered as the result of two opposite  phenomena, i.e.,  the diffraction and
the self-focusing caused
 by  dispersion and nonlinearity, respectively, for the \ivp\
of Eq.~\eqref{eq:nse}. Since their renormalization factor, $\varrho$, is a
function of both space and time, its use is not consistent with a quantum
mechanics analogy for the \nlse. As a consequence, the energy and force
densities obtained by Bian and Chan are somewhat artificial, since their filter
or renormalization factor affects the values of these
densities in a different manner depending on the soliton location at each
instant of time.

The 1-soliton solution of the initial-value problem for
Eq.~\eqref{eq:nse} is
    \begin{equation} \label{eq:1ss}
u(x,t)=A \sech ( \xi )
  \exp ( i \eta ).
    \end{equation}
where
    \begin{eqnarray} 
\xi &=& {A \over \sqrt{2}} ( x - x_0 - c t ) \\
\eta &=& {1 \over 2 } [  c (x-x_0) + (A^2 - {c^2 \over 2} ) t +  \phi_0],
    \end{eqnarray}
$A$, $c$, $x_0$ and $\phi_0$ are the soliton's amplitude, velocity,
initial position and initial phase, respectively, and the nonlinear
and diffraction force densities for $\varrho$=1 are, respectively,
    \begin{equation} 
f_n = -2 \sqrt{2} A^5 \sech^4 \xi \tanh \xi
    \end{equation}
    \begin{eqnarray} 
& &f_d = {A^3 \over 4} \sech^2 \xi \left[- 6 A c i \sech^2 \xi +
  8 \sqrt{2} A^2 \sech^2 \xi \tanh  \xi  + \mbox{} \right. \nonumber \\
 & &  \phantom {f_d = {A^3 \over 4} \sech^2 \xi  \left[\right.}
   \left. \mbox{} + \sqrt{2}( c^2 - 2 A^2 ) \tanh \xi  + 4 A c i \right]
    \end{eqnarray}
while the total force density, i.e., the sum of the nonlinear and diffraction
force densities, is
    \begin{equation} 
f_t =  {A^3 \over 4} \sech^2 \xi  \left[ - 6 A c i \sech^2   \xi +
  \sqrt{2}( c^2 - 2 A^2 ) \tanh \xi  + 4 A c i \right]
    \end{equation}

The above equations show that the  nonlinear force density is less than
zero behind the location of the \s 's maximum amplitude
and greater than zero in front  of it, whereas the real parts of both the
diffraction and total force densities depend on the values of $c$ and $A$.

The numerical  results of Bian and Chan~\cite{Bian91} indicate that  their total
force density, i.e., the sum of the real part of the diffraction force and
the nonlinear force  densities, vanishes for the 1-soliton solution, cf.
Eq.~\eqref{eq:1ss}.  For the \eNss~\cite{Zakh72,Gord83}, the energy
density  of Bian and Chan gives a nonvanishing force dominated by the nonlinear
density  force which indicates that the \s s are compressed while propagating.
However,  their interpretation can be criticized on the grounds that diffraction
effects on  the soliton must also be considered and that their renormalization
factor which is a
function of both space and time, affects in different manner the local force
densities.  Furthermore, it is important to note that their definition can cause
computational problems when calculating the nonlinear and diffraction force
densities  due the small denominator introduced by the filter in their
energy expression, especially in the \s\ tails.



\section{Boundary Conditions}

In this paper, the \nlse\ is studied numerically in a symmetric, finite
interval $\cal D$=$[-L,L]$ subject to the following homogeneous boundary
conditions
    \begin{equation} \label{eq:nse:r}
u(-L,t) + \gamma u_x(-L,t) = 0, \qquad u(L,t) + \gamma u_x(L,t) = 0
                    \text{\ and \ } \qquad t \geq 0.
    \end{equation}
The values $\gamma$=0 and $\infty$ correspond to Dirichlet and Neumann
boundary conditions, respectively.

Since the \nlse\ and the homogeneous Dirichlet and Neumann boundary
conditions are invariant under mirror reflections in $x$, their respective
initial-boundary-value problems also exhibit this invariance; as a consequence,
the interaction of a soliton with the left boundary is identical to that with
the right boundary. This symmetry or invariance is lost if $\gamma$ is
finite and different from zero, i.e., if homogeneous Robin boundary
conditions apply at both boundaries. In this paper, mixed boundary
conditions corresponding to $\gamma$=1 are considered, and the results for
the Dirichlet, Neumann and Robin boundary conditions are compared with
those for both the initial-value problem and  the
periodic boundary conditions
    \begin{equation} \label{eq:nse:p}
\ndpar{n}{u}{x}(x,t) = \ndpar{n}{u}{x}(x+2kL,t),
  \qquad \forall n \geq 0, \quad k \in \Z,
     \quad x \in {\cal D} \equiv [-L,L], \quad t \geq 0.
    \end{equation}


\section{Relationships between the Quantum Momentum and Energy Densities}

The most important invariants of the initial-value problem of the \nlse\ are
the first, second and third ones.
The first, known as wave mass or `number of particles', is
    \begin{equation} \label{eq:inv1}   
\int_{\cal{D} }  {|u|^2\,dx},
        \end{equation}
whose integrand will be denoted by
$\rho_{I1}$. The second invariant represents the total momentum in the
Hamiltonian  formalism and is given by
    \begin{equation} \label{eq:inv2}
\int_{\cal{D} }  {i \left  ( u^*u_x-uu_x^* \right) \,dx}
        \end{equation}
and whose integrand will be referred to as  $\rho_{I2}$. The third invariant is
the total energy or Hamiltonian, i.e.,
    \begin{equation} \label{eq:inv3}  
\int_{\cal{D} } {\left(|u_x|^2-\frac{1}{2} |u|^4\right) \,dx}
        \end{equation}
whose integrand is the Hamiltonian density of the \nlse\ and will
be denoted by $\rho_{I3}$.

The real and imaginary parts of the
quantum momentum density are
    \begin{equation} \label{eq:preal}
\re \{ p(x,t) \} = p_R = -\frac{i}{2} \left( u^*u_x-uu^*_x \right)
    \end{equation}
    \begin{equation} \label{eq:pimag}
\im \{ p(x,t) \} = p_I = -\frac{1}{2} \left( u^*u_x+uu^*_x \right)
    \end{equation}
those of the kinetic energy density
    \begin{equation} \label{eq:ekreal}
\re \{ e_k(x,t) \} = e_R = -\frac{1}{2} \left( u^*u_{xx}+uu^*_{xx} \right)
    \end{equation}
    \begin{equation} \label{eq:ekimag}
\im \{ e_k(x,t) \} = e_I = \frac{i}{2} \left( u^*u_{xx}-uu^*_{xx} \right)
    \end{equation}
and those of the diffraction force density
    \begin{equation} \label{eq:fdreal}
\re \{ f_d(x,t) \} = f_R = \frac{1}{2} \dpar{}{x} \left( u^*u_{xx}+uu^*_{xx}
\right)
    \end{equation}
    \begin{equation} \label{eq:fdimag}
\im \{ f_d(x,t) \} = f_I = -\frac{i}{2} \dpar{}{x} \left( u^*u_{xx}-uu^*_{xx}
\right)
    \end{equation}

From the above equations, the following  relations  may be easily
deduced
    \begin{equation} \label{eq:p1p2inv}
\dpar{}{t}\rho_{I1} = - 2  \dpar{p_R}{x} = \dpar{\rho_{I2}}{x}, \qquad
p_I = -\frac{1}{2} \dpar{}{x}\rho_{I1}.
    \end{equation}

It can also be easily shown using the \nlse\ that the following
equations hold
    \begin{equation} \label{eq:dp1dt}
\dpar{p_R}{t} = - \dpar{\rho_{I3}}{x} + f_R, \qquad \dpar{p_I}{t} = f_I,
    \end{equation}
    \begin{eqnarray} \label{eq:dp2dt}
\dpar{p}{t} = -  \dpar{\rho_{I3}}{x} -
\dpar{e_{k}}{x}, \qquad \dpar{\rho_{I3}}{t} =
\dpar{}{x}(u_{x}^*u_{t}+u_{x}u^*_{t})
    \end{eqnarray}

For initial-value problems, i.e., $\cal D$=$(-\infty,\infty)$  subject to $|u| \rightarrow
0$ as  $|x| \rightarrow \infty$, the above equations imply that
the total mass or number of
particles, the total momentum in the Hamiltonian formalism and the total energy
are invariant. Furthermore,
    \begin{equation}
\int_{\cal{D} } {p\,dx}=0,
    \end{equation}
i.e., the total quantum momentum is zero.

For quarter-plane or semi-infinite, initial-boundary-value problems,
i.e., $\cal D$=$[0,\infty)$  subject to $|u| \rightarrow 0$ as  $x
\rightarrow \infty$, the total number of particles is constant for
homogeneous Dirichlet or homogeneous Neumann boundary conditions at $x=0$, and
the imaginary part of the total quantum momentum is only zero for
homogeneous Dirichlet boundary conditions at $x=0$, whereas both the total
quantum momentum and its real part are different from zero.

For the finite-line, initial-boundary-value problems considered in Section
3, the total number of particles is constant for homogeneous Dirichlet or
homogeneous Neumann boundary conditions at both boundaries; the imaginary
part of the total quantum momentum is zero for periodic or for homogeneous
Dirichlet boundary conditions at both boundaries. The total
quantum momentum and its real part are different from zero.

\section{Numerical Method}

\label{se:numerical}

The Crank-Nicolson method was used to discretize Eq.~\eqref{eq:nse} in
the interval  $[-L,L]$, and a Newton-Raphson technique  was employed to solve
the resulting system of nonlinear  algebraic equations. The
diagonally dominant, tridiagonal system  of linear algebraic equations that
result from the Newton-Raphson  method was solved by means of a 2$\times$2
block-oriented version of the Thomas algorithm. In  the periodic case, a natural
optimization  of the Gaussian elimination technique for periodic problems that
yields  quasi-tridiagonal systems has been used before
applying the Thomas algorithm~\cite{Taha84}.

The Crank-Nicolson method is conservative since it preserves a discrete
equivalent of the first invariant, cf. Eq.~\eqref{eq:inv1}, i.e., the $L^2$ norm
of the  solution, for the discrete, initial-value problem, and for the two-point
initial-value problem with \P\ or \D\ \bc s.

The evaluation of the high-order derivatives of the soliton amplitude required
for the calculation of the soliton energy and force densities was performed  so
as  to preserve both  the maximum possible  symmetry in the finite difference
operators and  second-order accuracy. It should
be noted that, in the periodic case,   symmetric, second-order accurate in
space,  stencils were employed  at all the spatial grid points,
while, for the Dirichlet, Neumann and Robin  boundary conditions, asymmetric
stencils were introduced near the boundary  points in order to avoid the use
of  fictitious points.

\section{Presentation of Results}

In this section, some sample results (cf. Figs.~1--5) that illustrate the
nonlinear force density, the real part of the diffraction force density, the
total force density, i.e., the sum of the nonlinear force density and the real
part of the diffraction force density, and the real part of the momentum density
are presented as functions of space and time for the four types of boundary
conditions considered in this paper. Figures~1--5 also show the space-time
isocontours of the three-dimensional data presented in these figures. The
results presented in Figs.~1--5 as well as in Figs.~6--10 correspond to an
interval of $L$=50, $A$=$c$=1, $x_0$=$\phi_0$=0, and spatial and temporal step
sizes equal to 0.25 and 0.01, respectively.

Figure~1 corresponds to periodic boundary conditions and shows that the
nonlinear force density has  an $S$-shape which is nearly  the mirror reflection
of that for the diffraction force density. The total force density has also
an $S$-shape similar to the nonlinear one which indicates that, for periodic
boundary conditions, the nonlinear force density is larger than the diffraction
one. The momentum density has  a  bell-shape similar to that of the soliton
amplitude.

Figure~2 illustrates the large changes in the magnitude of the force densities
introduced by homogeneous Dirichlet boundary conditions and, especially, in the
collision of the soliton with the right boundary. It is interesting to note
that the shape of the nonlinear force density prior to and after the collision
of the soliton with the right boundary is the same owing to
the symmetry of the \nlse\ and boundary conditions; however, its magnitude
increases greatly in the collision process because the amplitude of the
soliton at the boundary is zero. Figure~2 also shows that the nonlinear force
density is larger than the diffraction one, and that the momentum density
changes sign once the soliton rebounds from the right boundary. Note that the
Dirichlet boundary conditions require that the momentum density at the boundary
be zero; therefore, the change in the momentum is smooth and the change in its
sign is associated with the different directions of the soliton velocity prior
to the collision and after the soliton rebounds from the boundary.

Figure~3 corresponds to homogeneous Neumann boundary conditions and illustrates
the $S$-shapes of the nonlinear, diffraction and total force densities prior to
and after the collision of the soliton with the right boundary; this collision is
accompanied by large increases in the magnitudes of these densities. During the
collision, the nonlinear force density shows a relative maximum near to the
right boundary, whereas the diffraction and total force densities exhibit
extrema. Figure~3 also indicates that, during the collision with the
right boundary, the diffraction force exceeds the magnitude of the nonlinear one
at the boundary, whereas the latter is larger than the former near to, but away
from, the boundary. The momentum density shown in Figure~3 changes sign upon the
collision of the soliton with the boundary, and its value at the boundary is
zero.

Comparisons amongst the momentum densities for the periodic, Dirichlet and
Neumann boundary conditions indicate that the soliton penetrates into the
boundary in the periodic and Neumann cases, i.e., $u(L,t) \ne 0$, whereas it does
not penetrate into the boundary for the Dirichlet case; the momentum preserves
its sign in the periodic case owing to the transparency of the boundary and has
the same shape for the Dirichlet and Neumann boundary conditions.

Since the \nlse\ with Robin boundary conditions is not a symmetric problem, the
interaction of a soliton with the left boundary is expected to be different from
that with the right one. For this reason, the soliton interactions with both
boundaries must be considered as shown in Figures~4 and~5 which correspond to
the first and second collisions, i.e., a collision with the right boundary
followed by another one with the left boundary. Figure~4 indicates
 the $S$-shapes of the nonlinear, diffraction and total force densities prior to
and after the collision of the soliton with the right boundary. The nonlinear
force density exhibits a relative maximum and a relative minimum near to, but
away from, the boundary where its value is not zero. Similar trends are observed
in the diffraction force density until its relative maximum becomes an extremum
at the boundary. Figure~4 also shows that the nonlinear force is greater than
the diffraction one except very near to the boundary during the collision
process. The momentum density presented in Figure~4 has the same shape as those
for the Dirichlet and Neumann boundary conditions, except that its value at the
right boundary is different from zero.

Figure~5 indicates the lack of symmetry of the \nlse\ subject to Robin boundary
conditions. In particular, the nonlinear force density maintains its $S$-shape
prior to and after the collision of the soliton with the left boundary; however,
compared with the collision with the right boundary, this density exhibits a
plateau before  reaching a negative extremum at the boundary. Figure~5  also
shows that the diffraction force  preserves its $S$-shape prior to and after the
collision of the soliton with the left boundary, and that it exhibits a plateau
near to and a positive extremum at the left boundary. The total force density
indicates that the nonlinear force is larger than the diffraction one, while the
momentum changes sign upon the collision of the soliton with the left boundary.
The isocontours presented in Figures 4 and 5 indicate that the soliton becomes
closer to the right boundary than to the left one.

In Figures 6--10, the negative values of the kinetic, potential and total
energy, and  of the momentum are presented as functions of time. Note that these
quantities are the integrals, from the left to the right boundary, of their
corresponding  densities, that the kinetic energy has been calculated from the
real part of its density, and that the momentum has been obtained from the
absolute value of its density.

The results presented in Figure~6 indicate that the kinetic and potential
energies are nearly constant and have opposite signs for the two-point,
periodic, initial-value problem. The total energy, i.e., the sum of the potential
and kinetic energies, is almost constant except for a minimum at about $t$=50
which corresponds to the time at which the soliton crosses the boundary. The
momentum presented in Figure~6 is also nearly constant.

Figure~7 corresponds to Dirichlet boundary conditions and shows that the kinetic
energy is larger than the potential one which is negative; the total
energy is positive and exhibits a maximum at about $t$=50 which corresponds to
the collision of the soliton with the right boundary. The results presented in
Figure~7 clearly indicate that the collision of the soliton with a Dirichlet
boundary is accompanied by an increase (decrease) in energy as the soliton
approaches (rebounds from) the boundary, and that
the energies after rebound are the same as those prior to the collision. The
momentum presented in Figure~7 indicates the deceleration (acceleration) of the
soliton as it approaches (recedes from) the right boundary; however, these
deceleration and acceleration processes do not cause large variations in the
soliton momentum.

The results presented in Figure~8 correspond to Neumann boundary conditions and
exhibit similar trends to those presented in Figure~7 except for the small,
relative minima that surround the relative maxima of the kinetic, potential and
total energies. Figure~8 also shows that the collision of the soliton with a
Neumann boundary produces larger increases in energy in smaller intervals of
time than that with a Dirichlet boundary. Large increases are also observed in
the momentum as the soliton collides with the right boundary. However, both the
energies and the momentum recover their values prior to the collision after the
soliton rebounding process is completed.

The energies presented in Figures~9 and~10 correspond to the first collision of
the soliton with the right and left boundaries, respectively, for the Robin
boundary conditions, and exhibit the same trends as, but have smaller magnitude
than, those corresponding to the Neumann boundary conditions presented in
Figure~8; the duration of the soliton interaction with the boundaries is shorter
(longer) than those of the Dirichlet (Neumann) boundary conditions presented in
Figure~7~(8). The most noteworthy features of the results presented in Figures~9
and~10 are the increase in the background radiation as the soliton collides with
the left boundary, the decrease (increase) in the momentum as the soliton
collides with the right (left) boundary, and the relative maxima and minimum
observed in the momentum upon the collision of the soliton with the left
boundary.


\section{Conclusions}

A quantum mechanics analogy is used to determine the forces on and the energies
of solitons governed by the \nlse\ subject to periodic and to homogeneous
Dirichlet, Neumann and Robin boundary conditions in finite intervals. For all
the boundary conditions considered in this paper, it has been shown that the
the nonlinear, diffraction and total force densities have $S$-shape profiles
and increase in magnitude as the soliton interacts with the boundaries. This
interaction is characterized by an increase in the force densities; the
largest increase corresponds to the homogeneous Neumann boundary conditions.

The values of the force and energy densities after the soliton rebound from the
boundary is completed, are same values as those prior to the
interaction with of the soliton with the boundary, and, except during the
collision process, the nonlinear force density is larger than the diffraction
one. During the collision process, the nonlinear force is larger than the
diffraction one for the Dirichlet boundary conditions; the former is smaller
than the latter at the boundary for the Neumann conditions.

The lack of symmetry of the Robin boundary conditions has been illustrated by
the different dynamics of the soliton collisions with the right and left
boundaries. In the collision with the right (left) boundary, the magnitude of
the diffraction (nonlinear) density force at the boundary  is larger (smaller)
than the nonlinear (diffraction) one, and the soliton becomes closer to the right
than to the left boundary. Since the nonlinear force is larger than the
diffraction one when the soliton is sufficiently far away from the boundaries,
it may be concluded that self-focusing effects are more important as regard the
quantum mechanics analogy developed in this paper.

The kinetic, potential and total energies of the soliton decrease as the
softness of the boundary conditions is increased, i.e., they are largest
(smallest) for Neumann (Dirichlet) boundary conditions, and are nearly constant
for the periodic boundary conditions.


\section*{Acknowledgments}

This research was supported by the Spanish D.G.I.C.Y.T. under Project  no.
PB91--0767. The second author (F.R.V.) has a fellowship from the
Programa Sectorial de Formaci\'on de Profesorado Universitario y Personal
Investigador, Subprograma de Formaci\'on de Investigadores "Promoci\'on General
del Conocimiento", from the Ministerio de Educaci\'on y Ciencia of Spain.



\newcommand{\bookref}[6]{#1, {\em{#2}}, #3 (#6).}

\newcommand{\paperref}[6]{#1, #2, {\em {#3}} {\bf{#4}}, #5 (#6).}

\newcommand{\paperrefiss}[7]{#1, #2, {\em {#3}} {\bf{#4}}(#7) #5 (#6).}

\newcommand{\procref}[7]{#1, #2, In {\em{#3}}, (Edited by #4), pp. #6, #5,
(#7).}

\newcommand{\procrefvol}[9]{#1, #2, In {\em {#3}}, #4, (Edited by #5), #6,
vol. #7, pp. #8 (#9).}

\newcommand{\procrefv}[8]{#1, #2, In {\em #3}, (Edited by #4), {\em #5} vol. #6,
pp. #7 (#8).}


\pagebreak

\clearpage

\section*{Figure Captions}

\begin{enumerate}

\item {Nonlinear (fn), diffraction (Re(fd)) and total (ftotal) force densities
and momentum density (p) for periodic boundary conditions.}

\item {Nonlinear (fn), diffraction (Re(fd)) and total (ftotal) force densities
and momentum density (p) for Dirichlet boundary conditions.}

\item {Nonlinear (fn), diffraction (Re(fd)) and total (ftotal) force densities
and momentum density (p) for Neumann boundary conditions.}

\item {Nonlinear (fn), diffraction (Re(fd)) and total (ftotal) force densities
and momentum density (p) for Robin boundary conditions and first collision
with the right boundary.}

\item {Nonlinear (fn), diffraction (Re(fd)) and total (ftotal) force densities
and momentum density (p) for Robin boundary conditions and first collision
with the left boundary.}

\item {Kinetic, potential and total energy, and momentum for periodic boundary
conditions.}

\item {Kinetic, potential and total energy, and momentum for Dirichlet
boundary conditions.}

\item {Kinetic, potential and total energy, and momentum for Neumann boundary
conditions.}

\item {Kinetic, potential and total energy, and momentum for Robin boundary
conditions and first collision with the right boundary.}

\item {Kinetic, potential and total energy, and momentum for Robin boundary
conditions and first collision with the left boundary.}

\end{enumerate}

\newpage
\thispagestyle{empty}
\begin{textblock*}{\paperwidth}(0mm,0mm)
   \noindent\includegraphics[width=\paperwidth]{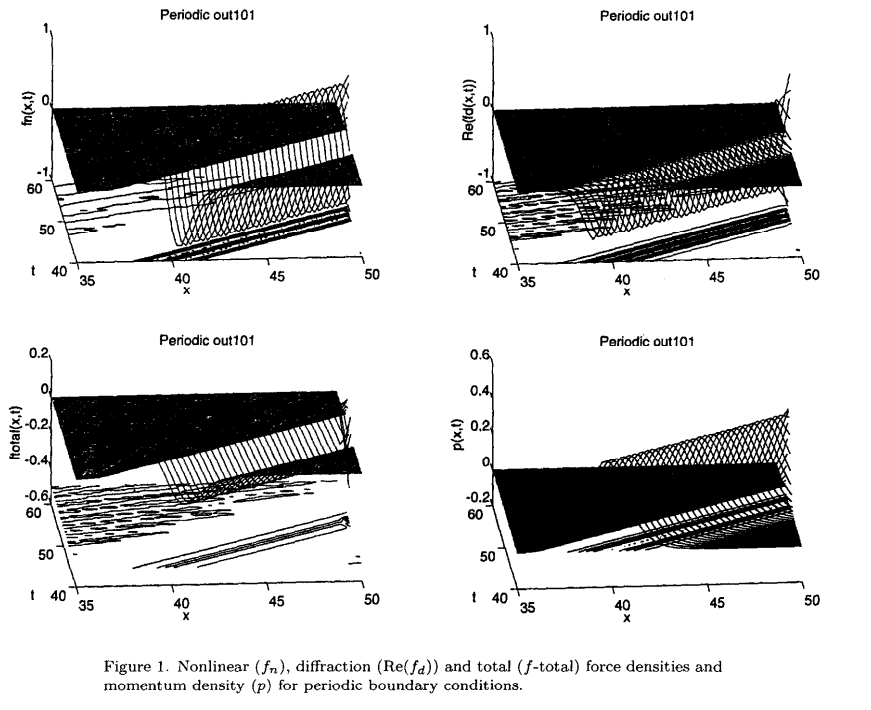}
\end{textblock*}
\mbox{}\newpage

\newpage
\thispagestyle{empty}
\begin{textblock*}{\paperwidth}(0mm,0mm)
   \noindent\includegraphics[width=\paperwidth]{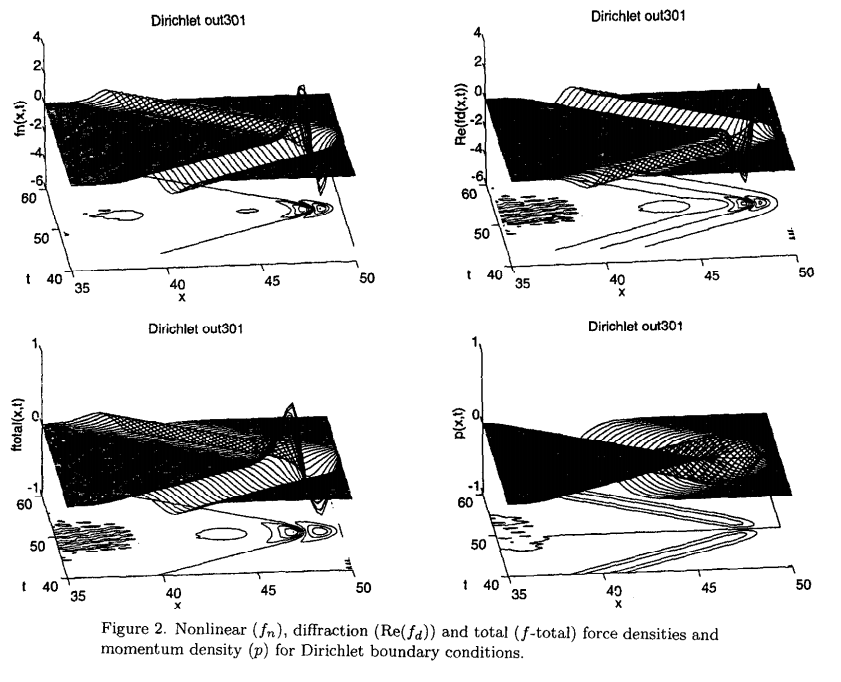}
\end{textblock*}
\mbox{}\newpage

\newpage
\thispagestyle{empty}
\begin{textblock*}{\paperwidth}(0mm,0mm)
   \noindent\includegraphics[width=\paperwidth]{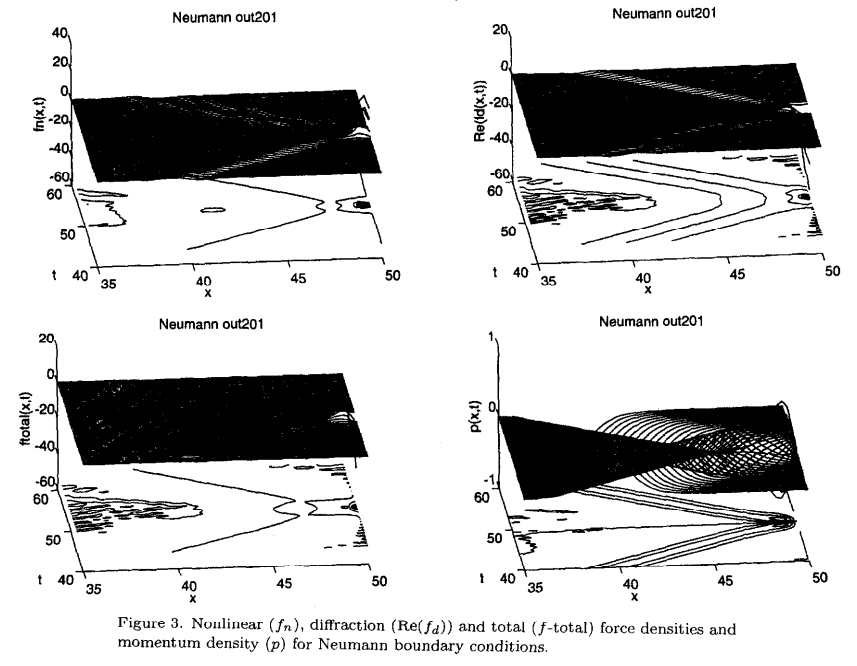}
\end{textblock*}
\mbox{}\newpage

\newpage
\thispagestyle{empty}
\begin{textblock*}{\paperwidth}(0mm,0mm)
   \noindent\includegraphics[width=\paperwidth]{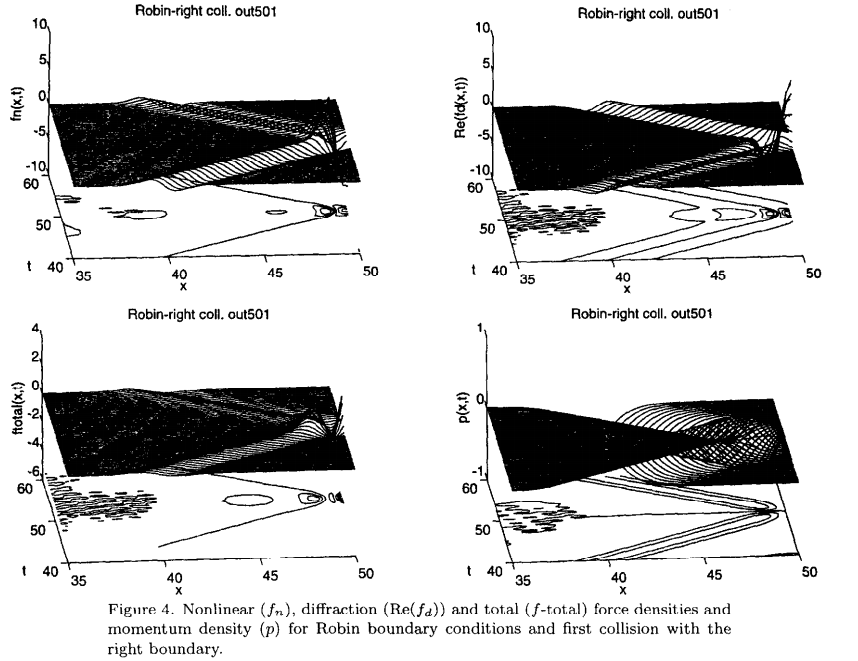}
\end{textblock*}
\mbox{}\newpage

\newpage
\thispagestyle{empty}
\begin{textblock*}{\paperwidth}(0mm,0mm)
   \noindent\includegraphics[width=\paperwidth]{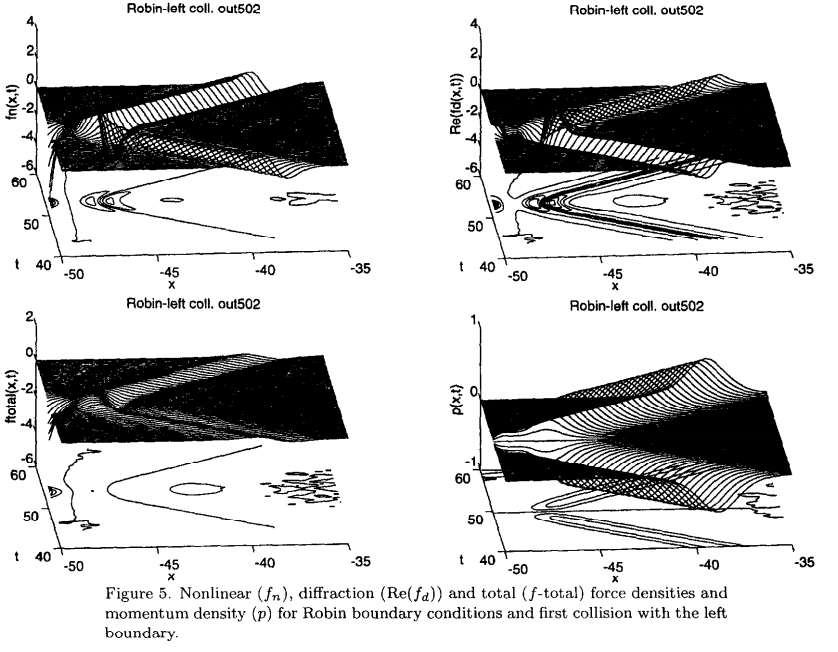}
\end{textblock*}
\mbox{}\newpage

\newpage
\thispagestyle{empty}
\begin{textblock*}{\paperwidth}(0mm,0mm)
   \noindent\includegraphics[width=\paperwidth]{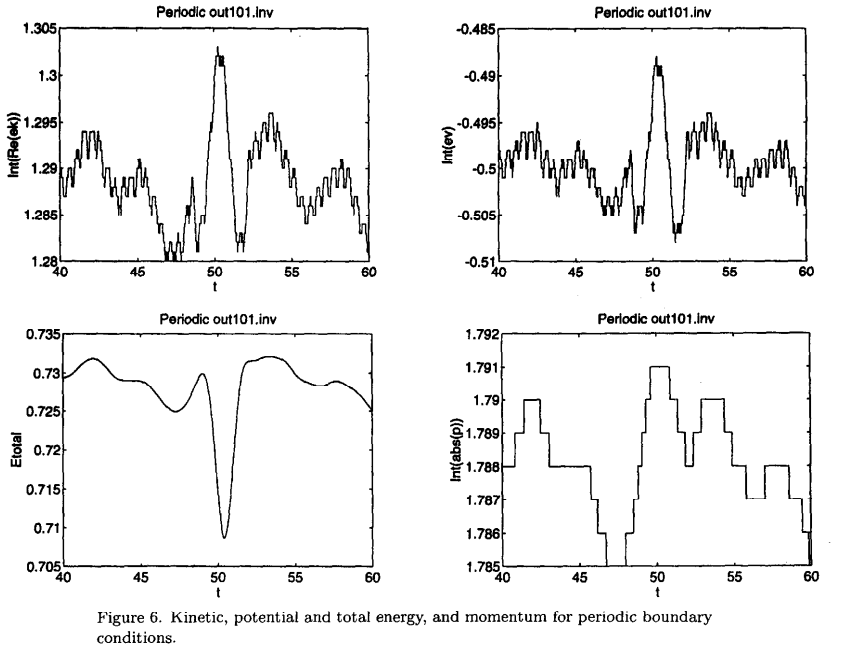}
\end{textblock*}
\mbox{}\newpage

\newpage
\thispagestyle{empty}
\begin{textblock*}{\paperwidth}(0mm,0mm)
   \noindent\includegraphics[width=\paperwidth]{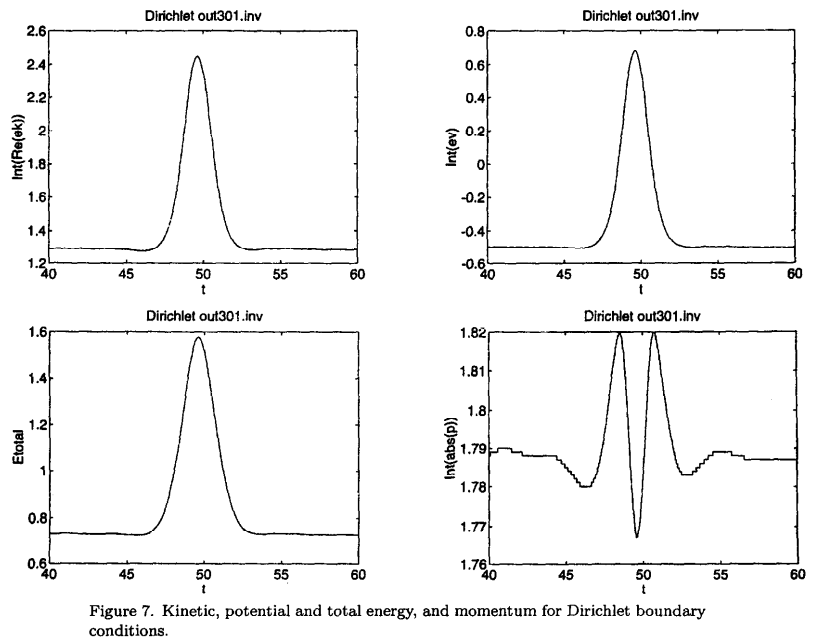}
\end{textblock*}
\mbox{}\newpage

\newpage
\thispagestyle{empty}
\begin{textblock*}{\paperwidth}(0mm,0mm)
   \noindent\includegraphics[width=\paperwidth]{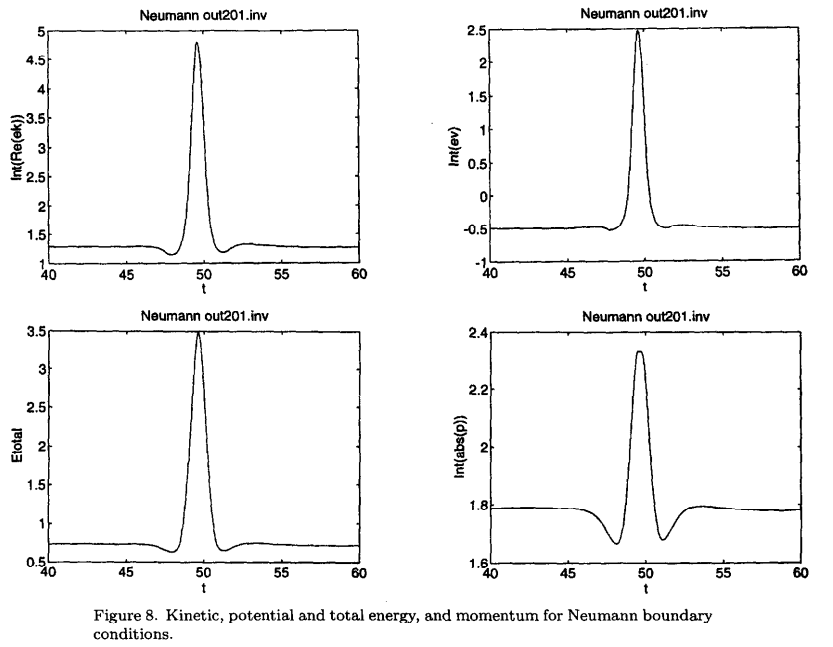}
\end{textblock*}
\mbox{}\newpage

\newpage
\thispagestyle{empty}
\begin{textblock*}{\paperwidth}(0mm,0mm)
   \noindent\includegraphics[width=\paperwidth]{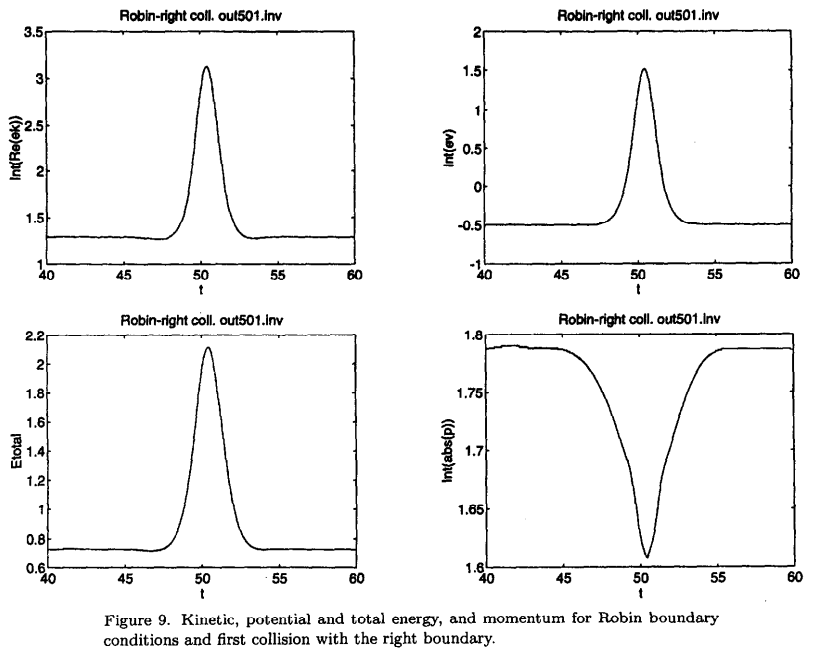}
\end{textblock*}
\mbox{}\newpage

\newpage
\thispagestyle{empty}
\begin{textblock*}{\paperwidth}(0mm,0mm)
   \noindent\includegraphics[width=\paperwidth]{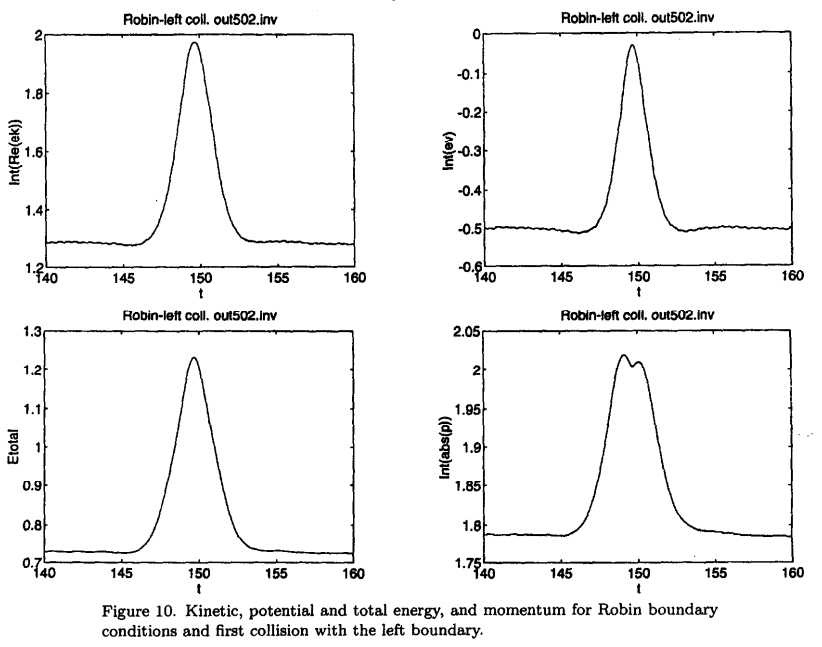}
\end{textblock*}
\mbox{}\newpage

\end{document}